
\magnification=1000
\baselineskip=16pt
\rightline{CERN-TH.6609/92}
\rightline{ILL-(TH)-92\# 16}
\vskip1truecm

\centerline{\bf SPECTROSCOPY, EQUATION OF STATE AND MONOPOLE}
\centerline{\bf PERCOLATION IN LATTICE QED  WITH TWO FLAVORS}
\vskip1truecm

\centerline {S. J. HANDS, A. KOCI\' C}
\centerline {\it Theory Division, CERN, CH-1211 Geneva 23, Switzerland}
\vskip3truemm

\centerline {J. B. KOGUT}
\centerline {\it Department of Physics, University of Illinois,
Urbana, IL  61801-3080, USA}
\vskip3truemm

\centerline {R. L. RENKEN}
\centerline {\it Physics Department, University of Central Florida,
Orlando, FL  32816, USA}
\vskip3truemm

\centerline {D. K. SINCLAIR}
\centerline {\it High Energy Physics Division,
Argonne National Laboratory, Argonne, IL  60439, USA}
\vskip3truemm

\centerline {K. C. WANG}
\centerline {\it Australian National University,
Canberra, A.C.T. 2600 Australia}
\vskip5truemm

\centerline {\bf Abstract}\footnote{$\,^{}$}{CERN-TH.6609/92}

Non-compact lattice QED with two flavors of light dynamical quarks is
simulated on $16^4$ lattices, and the chiral condensate, monopole density and
susceptibility and the meson masses are measured.  Data from relatively
high statistics runs at relatively small bare fermion masses of 0.005, 0.01,
0.02 and 0.03 (lattice units) are presented.  Three independent methods of
data analysis indicate that the critical point occurs at $\beta =0.225(5)$ and
that the monopole condensation and chiral symmetry breaking transitions are
coincident.  The monopole condensation data satisfies finite size scaling
hypotheses with critical indices compatible with four dimensional
percolation.  The best chiral equation of state fit produces critical
exponents ($\delta=2.31$, $\beta_{mag}=0.763$)
which deviate significantly from mean field expectations.  Data
for the ratio of the sigma to pion masses produces an estimate of the
critical index $\delta$
in good agreement with chiral condensate measurements.  In
the strong coupling phase the ratio of the meson masses are
$M_\sigma^2/M_\rho^2\approx 0.35$, $M_{A_1}^2/M_\rho^2\approx 1.4$
and $M_\pi^2/M_\rho^2\approx 0.0$,
while on the weak coupling side of the
transition $M_\pi^2/M_\rho^2\approx 1.0$,
$M_{A_1}^2/M_\rho^2\approx 1.0$,
indicating the restoration of chiral symmetry.\footnote{$\,^{}$}{August 1992}
\vfill\eject

\centerline {\bf 1.  Introduction and Review}
\vskip3truemm
This paper continues and sharpens our series of studies of noncompact QED
incorporating two species of light dynamical fermions.  The goal of this
paper will be to establish that monopole condensation in this theory is a
legitimate phase transition with power-law critical singularities, and that
this transition is coincident with chiral symmetry breaking.  The critical
properties of the chiral symmetry breaking transition will be studied using
the theory's meson spectrum and chiral condensate.  New methods of
analyzing the simulation data of the light quark sector of the theory will
suggest, as proposed in past work, that the critical point is characterized
by a non-trivial, interacting field theory.
        Although this paper will emphasize new simulation results and will
leave
parallel theoretical developments to other articles, we begin with some
general remarks and perspectives.  One of our motivations for this set of
studies is to understand whether realistic strongly coupled field theories
exist in four dimensions, and if so, to extract their essential
ingredients.  In the language of the renormalization groups, we are
searching for ultra-violet stable fixed points which are non-trivial.  We
accept the conventional wisdom that $\lambda\phi^4$
is trivial in four dimensions and
turn to theories with gauge fields.  QED is the simplest candidate in this
direction and if we let its charged fields be fermionic, chiral symmetry
can protect the model from unnatural fine-tuning problems.  An early
excursion in this field found that the compact lattice version of QED
coupled to dynamical fermions had a first order chiral symmetry breaking
phase transition at strong coupling, and was not a candidate for a
relativistic continuum quantum field theory [1]. The noncompact formulation of
the model was seen to have a second order chiral transition which became a
candidate for a continuum field theory [2]. However, since the transition
occurs at non-zero coupling conventional classical or
perturbatively-improved Actions do not necessarily yield reliable pictures
of the theory's critical point.  In fact, monopoles were discovered near
the chiral transition and large scale simulations suggested that a topological
 phase transition, monopole condensation, occurs
at the chiral transition [3]. This interesting result, which will be
strengthened considerably through additional simulations and finite size
scaling analyses here, means that conventional perturbative discussions of
the triviality of QED are not relevant here [4].  Theories with fundamental
monopoles are intrinsically nonperturbative because of the Dirac
quantization condition.  Studies of the dynamical nature of theories with
fundamental electric and magnetic charges [5]
are completely new even though the
subject is fifty years old [6]. Recent simulations have suggested that
the essential characteristics
of the critical point Action driving the physics we are observing in
simulations includes at least three ingredients:  1) a strong vector force;
2) induced four Fermi couplings; and 3) monopole condensation.  Item 2 has
been discussed and motivated elsewhere [7,8]
and will not be discussed explicitly
here.  Certainly, considering the richness of items 1-3, this particular
study is, alas, still in its infancy.  Nonetheless, the article will
present promising mass spectrum calculations which, when combined with
accurate chiral condensate measurements, will set the stage for decisive
critical exponent and anomalous dimension calculations which should help
clarify the theory's critical Action.

In Sec.2 we introduce a random refreshment procedure to make the hybrid
molecular dynamics algorithm more efficient.  In Sec.3 we shall present
data for the chiral condensate which suggests the existence of a
non-trivial phase transition at a coupling $\beta_c =0.225(5)$.  Since we have
previously done a similar study on a $10^4$ lattice, we shall monitor possible
finite size effects carefully.  In Sec.4 we study monopole condensation in
this theory and find that there is a phase transition at $\beta_c =0.225$ whose
singularities are compatible with four dimensional percolation.  In Sec.5
we consider the meson mass spectrum of the model.  A study of the ratio of
the pion and sigma masses is consistent with a power-law chiral phase
transition at $\beta_c =0.225(5)$
with critical indices agreeing with those found
in the Equation of State study of Sec.3.  An analysis of the dependence of
the pion mass on the chiral condensate also picks out $\beta_c =0.225$ and
suggests that the transition is non-trivial.  And finally, the systematics
of other meson mass ratios are studied and the level ordering of the theory
is presented.  In Sec.6 of this paper we present suggestions for further
study, both computational and analytic.

\vskip1truecm

\centerline
{\bf 2. A Random Refreshment Algorithm}
\vskip3truemm
The lattice Action and simulation algorithm have been presented and
discussed in great detail in past work.  The reader is referred to Ref.[9]
for explicit formulas.  As will be discussed below, the hybrid
molecular dynamics algorithm was considerably improved by introducing
"random refreshing".  This improvement tended to decorrelate low momentum
photon modes in our gauge configurations more efficiently than in past
studies.  To control systematic errors we were forced to use a small time
step in the molecular dynamics evolution equations.  In fact, on $16^4$
lattices we accumulated data with bare fermion masses m of 0.03, 0.02, 0.01
and 0.005 in lattice units.  The time step ranged from 0.02 at $m =0.03$ to
0.005 at $m =0.005$ to keep systematic errors smaller than statistical ones.
The criterion for the convergence of the conjugate gradient algorithm used
to invert the lattice Dirac operator was scaled with lattice size and
fermion mass to maintain uniform accuracy throughout our data set.  At each
coupling $\beta = 1/e^2$ and $m$ value
we generated between 200 and 400 molecular
dynamics trajectories.  Each trajectory was defined to be one time unit
long, so 400 trajectories at
$m =0.005$, where the time step was 0.005, corresponded to 80,000 sweeps of
the hybrid algorithm.  We collected data at couplings ranging from
$\beta = 0.15$
(strong coupling) to $\beta=0.30$ (weak coupling).  This study dwarfs all
others
in the literature.  Nonetheless, our meson spectrum data is not as accurate
as we would have liked.  In the last section of this paper we shall suggest
further improvements on the present work to generate more plentiful useful
results.

In performing our updating in the weakly coupled spin wave phase we
observed that the time development of the gauge field was very similar to
that for the free gauge field, although fermion feedback shifted the chiral
symmetry breaking transition to a considerably smaller value of the
coupling $\beta= 1/e^2$.  The normal refreshment scheme is inadequate for such
dynamics.  We have therefore developed a random refreshment scheme which is
adapted for the free field case and hence should be better than the
standard scheme for the problem at hand.  The free field lagrangian from
which we start is

$$
L=12\sum_{\rm links}\dot\Theta_\mu^2 -4\beta\sum_{{\rm sites},\mu,\nu}
\bigl(\Delta_\mu A_\nu - \Delta_\nu A_\mu \bigr)
\eqno(2.1)
$$
In momentum space, using the Feynman gauge for convenience, $L$ becomes

$$
L= \sum_{k=2\pi n/L}\biggl(
12\dot\Theta_\mu^2 (k) -
\beta\bigl(4-\sum_{\nu}\cos k_\nu\bigr)A_\mu^2(k)\biggr)
\eqno(2.2)
$$
i.e. a set of harmonic oscillators with frequencies

$$
\omega_k=\sqrt{2\beta\bigl(4-\sum_\nu \cos k_\nu\bigr)}
\eqno(2.3)
$$
so the periods $\tau_k$ of these oscillators lie in the range

$$
{\pi\over{2\sqrt{\beta}}}\leq\tau_k\leq
{ {2\pi}\over
{\sqrt{2\beta (1-\cos (2\pi /L))}}
}
\eqno(2.4)
$$
For $\beta= 0.225$ where the transition occurs on a $16^4$
lattice this range is 3.3 - 33.9.

Ideally for any one of these oscillators, one would like to refresh often
enough so that it samples its own (gaussian) energy distribution as rapidly
as possible and loses any phase coherence with the other oscillators.
However, we do not want to refresh it so often that it only random walks
through its own configuration space.  Our educated guess is that the best
choice would be a refreshment interval $\sim 1/\omega_k$ [10].
If we were to choose this
for the fastest oscillator, the slowest oscillator would random walk
through its configuration space, while if we chose this for the slowest
oscillator the fastest oscillator would spend most of its time retracing
its orbit.  We thus chose a compromise where the refreshment interval is
varied randomly to cover the "optimal" intervals for each of the
oscillators.  In fact we choose the refreshment interval each time to vary
between $\tau_{min}$ and $\tau_{max}$ according to

$$
\tau_{refresh}=\tau_{min}\bigl(\tau_{max}/\tau_{min}\bigr)^R
\eqno(2.5)
$$
where $R$ is chosen from a set of random numbers distributed uniformly over
the interval [0,1].  For our runs we chose $\tau_{min}=0.2$ and
$\tau_{max}= 10$.  For
runs somewhat longer than ours ($\sim 250$ time units per $\beta$)
we would suggest
increasing $\tau_{max}$ closer to the maximum period of the oscillators.

\vskip1truecm

\centerline {\bf 3. Chiral Equation of State}
\vskip3truemm
The data for the average plaquette and $<\bar\psi\psi>$
at various couplings $\beta =
1/e^2$ and bare fermion masses $m$ are presented in Tables 1 and 2,
respectively.  The $<\bar\psi\psi>$ data is plotted in Fig. 1.  A particularly
useful, relatively unbiased way to extract physics from these measurements
it to construct the theory's equation of state [11],

$$
{ {<\bar\psi\psi> }\over{m^{1/\delta}} }=
F\biggl( { {\Delta\beta}\over{<\bar\psi\psi>^{1/\beta_{mag}}} } \biggr)
\eqno(3.1)
$$
where $\Delta\beta =\beta-\beta_c$, $\beta_c$
is the bulk critical point for chiral symmetry
restoration, and $\delta$ and $\beta_{mag}$
are critical exponents.  At $\beta=\beta_c$, $\delta$  records
the response of the order parameter to a small symmetry breaking field,

$$
<\bar\psi\psi> = C m^{1/\delta}
\eqno(3.2)
$$
while $\beta_{mag}$ controls the shape of the $<\bar\psi\psi>$
vs. $\beta$ curve in the broken phase in the chiral limit $m\to 0$,

$$
<\bar\psi\psi> =D(\beta_c -\beta)^{\beta_{mag}}
\eqno(3.3)
$$
Eq.(3.3) is difficult to use in practice because it refers to
$<\bar\psi\psi>$
at vanishing fermion mass $m$, a quantity which is not directly accessible
to our algorithm.  Our strategy will then be to exploit Eq. (3.2) to find
candidates for $\beta_c$ and $\delta$,
and then construct the equation of state Eq.(3.1)
for various $\beta_{mag}$.
Only for the correct, physical values of $\beta_{mag}$ will all
the data of Table 2 fall on the universal curve
Eq. (3.1).  Past studies on smaller lattices and polynomial extrapolations of
$<\bar\psi\psi> $ to zero $m$ at fixed
coupling $\beta$, indicate that the critical coupling is near 0.225 and
$\beta_{mag}$ is
between 0.50 and 1.00 [3]. To search for the power-law behavior of Eq. (3.2)
expected at the critical point, we plot $-\ln <\bar\psi\psi> $
vs. $-\ln m$ for $m=
0.005$, 0.01 and 0.02 in Fig. 2.
Note that at $\beta=0.22$, the $<\bar\psi\psi> $ value at
$m =0.005$ is larger than the power-law fit to the $m=0.01$ and 0.02 data
points, while at $\beta=0.225$ it is smaller than the fit.  This suggests that
$\beta=0.220$ lies in the broken symmetry phase and the critical coupling is
between 0.220 and 0.225.  The straight line fits in Fig. 2, then predicts
that the critical index $\delta$ lies between 2.31 and 2.75.
The reader might recall
that this procedure was extremely successful for the quenched model where
much greater statistics and smaller quark masses could be probed [12]. In that
case data was taken with a resolution of 0.001 in $\beta$ rather than 0.005
used
here and the critical coupling was determined to three significant digits.

It is also interesting to search for $\beta_c$ and $\delta$
in a slightly different way.
Consider the plots of $-1/\ln (m)$ vs. $-1/\ln <\bar\psi\psi>$
at fixed $\beta$ in Fig.3.  This fit is particularly suitable for determining
the position of the critical coupling and the value of the critical exponent
$\delta$. Unlike the previous plot, it brings the chiral limit,
$m=0$, to the origin. Consider the EOS in its standard form [11],
$m=<\bar\psi\psi>^\delta f(z)$,
with $z$ being the reduced variable as defined by
Eq.(3.1). The universal function $f(z)$
is monotonically decreasing with only one zero
in the broken phase. Denote $x=-1/\ln <\bar\psi\psi>$ and
$y=-1/\ln (m)$. Then,

$$
y= {x\over{\delta-x\ln f(z)} }
\eqno(3.4a)
$$
Because $f(z)$ vanishes in the chiral limit in the broken phase, the curve
$y=y(x)$ approaches the horizontal axis
with an infinite slope, while in the symmetric
phase it approaches the origin linearly, with unit slope.
At the critical point $z=0$ we have,

$$
y_c= {x\over{\delta-x\ln f(0)} }
\eqno(3.4b)
$$
As we will see in later sections, the
universal function $f(z)$ will turn out
to satisfy $f(0)\approx 1$ so that the second term in the numerator of
Eq.(3.4b)
can be neglected and the slope of the curve $y_c\approx x/\delta$ gives a
fairly good estimate of the exponent $\delta$.
{}From these two equations it follows that

$$
{1\over{y_c}}- {1\over y}= \ln { {f(z)}\over{f(0)} }
\eqno(3.5)
$$
Considering the monotonicity of the function $f(z)$, the difference
$y(x)-y_c(x)$ changes sign as the critical point is crossed. All the
curves in the broken phase lie below, and those in the symmetric phase lie
above the critical line. The lines lying in the
broken phase terminate on the horizontal axis, whereas those
in the symmetric phase and at the critical point terminate at the origin.
Except at the critical point, all the lines have curvature
downwards. From the figure we see that the straight line hypothesis
works well at $\beta=0.225$ where it gives $\delta=2.29$ in good
agreement with Fig. 2.  In addition, the curves for $\beta=0.22$ and 0.210
suggest that $<\bar\psi\psi>$ is non-zero in the chiral limit
at $m\to 0$, while the
curves for $\beta=0.23,0.24$ and 0.26 indicate that
$<\bar\psi\psi>$ vanishes in the chiral limit.

Consider the equation of state Eq.(3.1) for different hypothetical $\beta_c$
values 0.230, 0.225 and 0.220.  For each $\beta_c$ and $\delta$
value we can inspect Eq.(3.1) as a function of $\beta_{mag}$.
We will show just three characteristic plots
from the multitude we viewed.  Recall the definition of the susceptibility
critical index $\gamma$, $\beta_{mag} = \gamma /(\delta-1)$.
We shall show numerical evidence for the
equality $\gamma =1$ later in this article.  Recall also that $\gamma$
is exactly unity
in the quenched model as well as in mean field theories of chiral
transitions.  So, we will investigate the hypothesis $\gamma = 1$ and show in
Figs. 4, 5 and 6, the chiral equation of state for the three cases:  (1)
$\beta_c =0.220$, $\delta = 2.75$, $\beta_{mag} =0.571$, (2) $\beta_c =
0.225$, $\delta =2.31$, $\beta_{mag} = 0.763$, and
(3) $\beta_c =0.230$, $\delta = 1.89$, $\beta_{mag} = 1.124$.
The universal character of the
plots -- that all the data at various couplings and quark masses fall on one
curve -- is fairly good.  Inspecting the three figures, we see that the
dispersion in the $\beta_c =0.225$ case is the
least, and this fact suggests that

$$
\beta_c =0.225, \,\,\,\,\,\,\,\,
\delta = 2.31,  \,\,\,\,\,\,\,\,
\beta_{mag}=0.763
\eqno(3.6)
$$
as our ``preferred'' conclusion from these measurements.

These $16^4$ results should be contrasted with our $10^4$ results published
earlier [3].  Comparing the raw
$<\bar\psi\psi>$ data between the two lattice
sizes, we note considerably more precision in our present study which
accounts for the fact that we have stronger conclusions here.  Our
preferred values for $\beta_c,\delta$ and $\beta_{mag}$
as listed in Eq.(3.6) are compatible
with our $10^4$ results.  The only significant finite size effects in our
chiral condensate measurements occur at the smallest $m = 0.005$ value where
the $<\bar\psi\psi> $
values on the larger lattice lie above those on the smaller
lattice.  This is the expected systematic trend - the smaller lattice acts
as a ``finite temperature'' environment, suppressing chiral symmetry
breaking.  Note that at $\beta_c =0.225$, $<\bar\psi\psi> =0.0881(29)$
on a $10^4$ lattice
and $<\bar\psi\psi> = 0.0943(8)$ on a $16^4$ lattice which is a $6\%$
shift ignoring error bars and a two standard deviation effect otherwise.

We shall see in Sec.5.b below that meson spectroscopy also favors $\beta_c=
0.225$ and $\delta = 2.31$ in a completely independent fashion.
\vskip1truecm

\centerline {\bf 4.  Monopole Condensation}
\vskip3truemm
In past work we have identified $\beta_c =0.225$
as the critical point for monopole condensation,
using order parameter and susceptibilities first suggested in Ref.[13].
We can now sharpen that analysis using finite size scaling analyses
to argue that monopole condensation is a true phase transition with
power-law thermo-dynamic singularities.  The precise definition of bond
percolation and the identification of monopoles within a gauge field
configuration can be found in Ref.[13] and will not be repeated here.  Our
order parameter for the monopole condensation transition will be
$M=n_{max}/n_{tot}$
which is the ratio of the number of dual sites in the largest
monopole cluster to the total number of connected sites in all monopole
clusters.  Fluctuations in $M$ generate a monopole susceptibility
$\chi$ which
should diverge at the critical point.  In Fig.7 we show
$\chi$ vs. $\beta$ on the $16^4$
lattice for bare quark masses of 0.03, 0.02 and 0.01.  There is no systematic
dependence on the quark mass over this range of relatively small values.
The critical point for monopole condensation is seen to be
$\beta_c = 0.225(5)$.
Recall that in the quenched noncompact lattice QED model $\chi$ peaked at
weaker
coupling $\beta=0.244$, so charge screening has pushed the monopole
condensation
point to stronger coupling as expected [14].  However, the peak values of
$\chi$
are within error bars in the two cases.  Finite size scaling relates the
maximum value of $\chi$, $\chi_{max}(L)$, on a lattice of volume
$L^4$ to the critical
indices $\gamma_{mon}$ and $\nu_{mon}$ of the monopole condensation transition,

$$
\chi_{max}\sim L^{ \gamma_{mon}/\nu_{mon} }
\eqno(4.1)
$$
In Fig.8 we show
$\ln \chi_{max}$ plotted against $\ln L$ for $\beta_c =0.225$ for the data
presented in Table 3.  For each lattice size we confirmed that
$\beta_c = 0.225(5)$
- no size dependence was observed in the critical point of this transition
(This peculiar fact was also observed in the quenched case [14].).  Note that
a linear fit to Fig.8 is very good and we read off the ratio of the
critical indices $\gamma_{mon}$ and $\nu_{mon}$,

$$
\gamma_{mon}/\nu_{mon} =2.16(6)
\eqno(4.2)
$$
In the quenched case we measured $\gamma_{mon}/\nu_{mon} = 2.24(2)$.

Now consider the behavior of the order parameter $M = n_{max}/n_{tot}$.  First
consider histograms of $M$ on the $16^4$ lattice at $m=0.01$
and variable $\beta$.  In
Fig.9a we consider weak coupling $\beta = 0.24$ and find that $M$ is
essentially
zero with little dispersion.  However, at $\beta = 0.23$ shown in Fig.9b, the
distribution has moved to positive $M$ and has broadened.  At $\beta = 0.225$
shown
in Fig.9c the effect is most dramatic.  Continuing to $\beta = 0.22$
(Fig.9d), $\beta
=0.21$ (Fig.9e) and $\beta=0.20$ (Fig.9f), we see the order parameter increase
in value and its fluctuations diminish.  A plot of $M$ vs. $\beta$ will also be
presented below.

A particularly effective way to measure the "magnetic" critical exponent
$\beta_{mon} $,

$$
M(\beta) = C (\beta_c -\beta)^{\beta_{mon} }, \,\,\,\,\,\,\,\,
(\beta<\beta_c)
\eqno(4.3)
$$
which controls the nonanalytic vanishing of $M$ in the $L$ limit, is to use
finite size scaling.  At the critical point and on a $L^4$ lattice
$M(\beta_c,L)$ should vanish with a power-law,

$$
M(\beta_c,L) \sim L^{-\beta_{mon}/\nu_{mon}},
\eqno(4.4)
$$
Using the data in Table 3 we show $\ln M(\beta_c,L)$ vs. $\ln L $
in Fig.10 for $L$
ranging from 10 to 18.  The error bars are much larger than we would
prefer, but an estimate of $\beta_{mon}/\nu_{mon}$, follows from the
$L = 12, 14, 16$ and 18 points:

$$
\beta_{mon}/\nu_{mon}=0.64(22)
\eqno(4.5)
$$
Similar measurements in the quenched model [14] gave
$\beta_{mon}/\nu_{mon}=0.88(2)$.
The values of $M(\beta_c,L) $ for the quenched model were systematically below
those of Table 3.  It is not clear whether fermion feedback has reduced
$\beta_{mon}/\nu_{mon}$,
below its quenched value since the error bars in Fig.10 are
relatively large.

In Ref.[14] we argued that monopole condensation in quenched noncompact
lattice QED is a percolation process uncomplicated by dynamical
considerations.  Our perspective on this problem comes from bond
percolation models where one randomly occupies bonds with a probability
$\rho$.
If $\rho$ is small, typical lattice configurations
have small, improbable connected clusters
of occupied bonds.  However, as $\rho$ increases towards a critical
concentration $\rho_c$, the largest connected cluster becomes infinite in
extent
and occupies a macroscopic fraction of the available lattice bonds.  The
value $\rho_c$ is called the percolation threshold and traditional statistical
mechanics calculations produce the estimate $\rho_c  \approx 0.16$ in four
dimensions.  Following Ref.[13] we computed the total amount of monopole
current $\ell$, $\ell =\sum_{n,\mu} |m_\mu(\tilde n)|$,
where $m_\mu(\tilde n)$ is the
integer-valued monopole current defined on links of the dual lattice, in
each of our $16^4$ configurations.  The resulting curve of the average value
of $\ell$ normalized by the number of lattice links
$N_{\ell}$ is shown in Fig.11.  We
also show the percolation order parameter
$M = n_{max}/n_{tot}$ there for the same
set of configurations having small
$m = 0.01$.  We see that $M$ turns on at $\beta =0.225$ with the expected
finite
size rounding and that
$<\ell>/N_\ell$ is  roughly 0.135 and varies smoothly around
the critical point.  We learn that $<\ell>/N_\ell$ is near the critical
concentration of pure bond percolation and since it varies smoothly it
could be used as an independent variable to parameterize the monopole
measurements instead of the gauge couplings $\beta$.  It is interesting that
$<\ell>/N_\ell$
varies abruptly across the confinement transition of pure compact
U(1) lattice gauge theory [15].  The character of Fig.11 suggests that the
monopole dynamics of the two flavor noncompact model is not qualitatively
different from the quenched model where accurate finite size scaling
measurements placed the monopole transition in the universality class of
four dimensional percolation [14].

In summary, the two flavor theory has a monopole condensation transition
at $\beta_c =0.225 $
with power-law thermodynamic singularities.  The determination
of $\gamma_{mon}/\nu_{mon}$
is in good agreement with the same calculation for the
quenched model which in turn has critical indices in agreement with four
dimensional percolation [16]. The measurement of $\beta_{mon}/\nu_{mon}$  is
 less
quantitative, but the data for $M = n_{max}/n_{tot}$
certainly satisfies a scaling
hypothesis.  It is tempting to conclude that the chiral symmetry breaking
transition and the monopole condensation transition are coincident.  If
that is true, then the set of field theory critical indices controlling the
chiral transition $(\delta = 2.31$, $\beta_{mag} =0.763) $
would be distinctly different from mean field values.

It is sometimes said that the monopoles of lattice QED are artifacts
of the regularization scheme
and have no physical significance. However, in this case we have seen that
they experience a second order phase transition with scaling laws and
critical indices which satisfy hyperscaling. Therefore, their critical
properties have decoupled from the lattice itself and they should have
macroscopic physical consequences.
\vskip1truecm

\centerline{\bf 5.  Spectrum}
\vskip3truemm

\noindent a.  Raw Data

We calculated the propagators of the pion, sigma, rho, and $A_1$ mesons using
standard techniques borrowed from lattice QCD large scale simulations [17].
Most of the meson propagator calculations used a point source for the quark
propagator calculation and then took the zero momentum projection of the
propagator perpendicular to the ``temporal'' direction of the $16^4$ lattice.
The fall-off of the meson propagator in the temporal direction then
determined the mass of the state using fitting technology borrowed from
lattice QCD [17].  The standard parameterization reads,

$$
G(t) =  \sum_i C_i \biggl(\exp \bigl(-M_i\tau\bigr) +
\exp \bigl(-M_i(16-\tau)\bigr) \biggr)+
\sum_i (-1)^\tau D_i  \biggl(\exp \bigl(- \tilde M_i \tau\bigr)
+ \exp\bigl(-\tilde M_i (16-\tau)\bigr)\biggr)
\eqno(5.1)
$$
and we considered fits over different ranges of $\tau$ between 0 and 8.  In
most
cases good fits were found with just the leading single particle term in
Eq.(5.1) determining the amplitude $C$, and the mass $M$, but in several cases
secondary particles were also necessary although their masses and
amplitudes were not determined with sufficient accuracy to be interesting.
Error estimates on the masses followed from the usual correlated $\chi^2$
procedures, also borrowed from lattice QCD studies [17].

Before turning to the masses themselves, it is instructive to consider
typical fits.  Consider the coupling $\beta=0.225$ and $m = 0.01$, and the pion
(Fig.12), sigma (Fig.13), rho (Fig.14) and $A_1$ (Fig.15) propagators.  We
show both the raw data (open circles) and the fits (triangles).  As usual,
the fits for the pion were the best, and we will be able to do some
quantitative
 analysis with the data here.
 The sigma particle also admitted respectable fits and its mass estimates
will be useful.  The rho and $A_1$ were relatively disappointing and very
little quantitative information will be obtained in these cases.

The data of meson masses (squared) for couplings ranging from 0.20 to 0.26
and quark masses from 0.03 to 0.005 are given in Tables 4, 5, 6 and 7.  The
quality of the fits is indicated in part by the error bars accompanying
each entry.  Typically, especially in the broken symmetry phase, the
confidence levels of the fits were very good ($90\%$ and higher confidence
levels were common).  However, the quality of the fits in the weak coupling
phase was distinctly poorer.  The simple hypothesis Eq.(5.1) is probably
too naive in the weak coupling phase where light charged fermions propagate
freely, interacting with massless photons, etc.  This result is certainly
not very surprising, but we did not know better fitting forms than
Eq.(5.1) which could be used productively on our limited data set.  In
addition, examining the rho and $A_1$ tables, we see that the large
statistical errors in the data renders the fits of little use.  Missing
entries in the tables indicate that the fitting routine was unable to find
an acceptable fit for that particular propagator.  Clearly it is
meaningless to extrapolate the rho and $A_1$ masses to $m\to 0$.  Therefore,
our
analysis will consider $m =0.01$ for the most part for these two states.
Happily, the pion and sigma results are much more quantitative and we will
learn a lot from their $m$ and $\beta$ dependencies.
\vskip5truemm

\noindent b.  The Goldstone Pion

If chiral symmetry is really broken at $\beta_c =0.225$, then the pion should
be
massless in the strong coupling phase.  This effect is quite clear in the
data.  If we extrapolate the pion mass-squared to $m= 0$, zero symmetry
breaking, then Fig.16 follows.  The $m\to 0$ extrapolation was done using the
$m=0.03$, 0.02 and 0.01 data and the form

$$
M_\pi^2= a + b\, m + c\, m^2 + ...
\eqno(5.2)
$$
expected from PCAC.  Although the error bars in Fig.16 are quite large,
$M_\pi^2$ is compatible with zero for $\beta\leq 0.225$ and is non-zero in the
weak coupling phase.  Recall, however, our caveat about the fitting
procedure Eq.(5.1) at weak coupling   the simple pole structure of
composite meson propagators in momentum space is probably too naive in the
weak coupling phase to parameterize the physics quantitatively.
Nonetheless, the qualitative fact that the pion mass is distinctly non-zero
in the weakly coupled phase is confirmed by the simulation.  Chiral
symmetry restoration in the weak coupling phase will be discussed in
greater detail below.
\vskip5truemm

\noindent c.  Spectroscopy and the Critical Index $\delta$

We have argued in Ref.[18] that the properties of ratios of dynamically
generated masses $M_a$ and $M_b$, $R_{ab}(m,\beta)=
G(m/(\beta-\beta_c)^\Delta))$, with $\Delta=\delta\beta_{mag}$,
are determined by the Equation of
State of the chiral order parameter.  Let $R(m,\beta)$ denote the ratio of
$M_\pi^2$ to $M_\sigma^2$ and write the Equation of State in the standard form,

$$
m = <\bar\psi\psi>^\delta f\biggl(
{ {\Delta\beta}\over{ <\bar\psi\psi>^{1/\beta_{mag}}} }\biggr)
\eqno(5.3)
$$
with $\Delta\beta= \beta_c - \beta$.  Then [18],

$$
{1\over{R(m,\beta)}} = \delta -   {x\over{\beta_{mag}}}
{ {f'(x)}\over{f(x)}}
\eqno(5.4a)
$$
where $x$ is the reduced coupling,
$\Delta\beta/<\bar\psi\psi>^{1/\beta_{mag}}$.  Eq.(5.4a) reduces
at the critical point $\beta = \beta_c$ $(x = 0)$ to,

$$
R(m,\beta_c)  = { {M_\pi^2}\over{M_\sigma^2}}={1\over\delta}
\eqno(5.4b)
$$
This result will give us another measurement of $\beta_c$ and
$\delta$ to supplement and,
in fact, support Eq. (3.6).  Recall that the derivation of Eq. (5.4a-b)
follows from the chiral equation of state describing the spontaneous
symmetry breaking of a continuous group [18].  In fact, the derivation
reveals more than Eq. (5.4b) which states that at the critical point the
ratio of the pion and sigma masses squared is independent of $m$ and is
equal to $1/\delta$.  For fixed $\beta<\beta_c$
in the broken symmetry phase, $R(m,\beta)$ should
fall to zero as $m\to 0$ since the pion is a Goldstone boson.  For fixed
$\beta>\beta_c$
in the symmetric phase, $R(m,\beta)$ should increase to unity as $m\to 0$
since the
sigma and pion must be degenerate in the symmetric phase. Thus, in the scaling
region, the following bound can be placed on $\delta$
[18]\footnote{$\,^\dagger$}{
An analysis very similar to ours was used to determine the universality
class of the finite temperature
chiral restoration transition in QCD at infinite couplings by,
G. Boyd, J Fingberg, F. Karsch, L K\" arkk\" ainen and B. Peterson,
Nucl. Phys. {\bf B376}, 199 (1992).}

$$
R(m,\beta\leq\beta_c) \leq {1\over\delta}\leq R(m,\beta\geq\beta_c).
\eqno(5.5)
$$
Because of the scaling form of the mass ratio, this bound improves as
either $\beta$ approaches $\beta_c$ or as $m$ becomes large.

In Fig.17 we plot $R(m,\beta)$ using the data for $M_\pi^2$ and
$M_\sigma^2$ in
Tables 4 and 5.  For $\beta =0.20$, 0.21 and 0.22 $R$
clearly falls as $m$ decreases
while for $\beta = 0.23$, 0.24 and 0.25 $R$ increases as
$m$ decreases.  Only at $\beta =
0.225$ is $R$ independent of $m$ and has the value 0.391(11) which gives

$$
\delta =2.56(7)
\eqno(5.6)
$$
The consistency of these results with the analysis based on the chiral
equation of state in Sec.3 is quite striking.  In fact, the determination
of $\beta_c =0.225(5)$ is most persuasive from Fig.17 --
$R(m,\beta)$ systematically
decreases with $m$ at $\beta =0.22$ and systematically increases as
$m$ decreases
at $\beta = 0.23$, thus picking out $\beta_c = 0.225(5) $ as the critical
point.

It is interesting to pursue Eq. (5.3) and (5.4) in somewhat greater
detail.  In particular, Eq. (5.4a) states that the mass ratio $R =
M_\pi^2/M_\sigma^2$ follows from the Equation of State with no additional
parameters, etc.  Do our chiral condensate measurements produce a universal
Equation of State $f(x)$ which predicts the correct systematics for $R$?  This
is a highly nontrivial test of the hypothesis that there is a chiral
critical point at  $\beta_c = 0.225 $ with $\delta = 2.31 $
and $\beta_{mag} = 0.763 $ since the
spectroscopy measurements of $M_\pi^2$ and $M_\sigma^2$  were independent of
chiral condensate considerations.  In Fig.18 we plot the Equation of State
in the "standard" form Eq. (5.3).  One of the advantages of plotting the
Equation of State in this form is that the universal function $f$  is so
simple.  Clearly a linear fit for $f$ represents the data adequately.  In
fact, as discussed in Ref.[18] a linear $f$ occurs only in a theory where
the susceptibility index $\gamma$ is unity.  As mentioned in our discussion of
Fig. 4-6 and now seen in Fig.18, the simulation data does provide direct
evidence for $\gamma = 1$.  The interested reader is directed to Ref.[18] for
further detail, since here our emphasis is on spectroscopy and the exponent
$\delta$.  A linear fit to $f(x)$ gives,

$$
f(x) = -5.3125  x  + 1.075
\eqno(5.7)
$$
Using Eq.(5.4a) $R(m,\beta)$ is now determined in a universal form and is
plotted in Fig.19.  Comparing Fig.17 and Fig.19 we find good agreement
within the rather substantial error bars for our spectroscopy measurements.
Nonetheless, the general success of Eq.(5.3) and (5.4) to describe
spectroscopy data within the hypothesis of a chiral critical point with
power-law non-mean field singularities is encouraging.
\vskip5truemm

\noindent
d. $M_\pi^2 $ vs. $<\bar\psi\psi>^2$, the Critical
Point and Anomalous Dimensions

As we have discussed in Ref.[18], measurements of the pion mass and the
chiral order parameter can be exploited in unison to locate critical points
and measure anomalous dimensions.  The two theoretical ingredients
underlying this strategy is 1) correlation scaling; and 2) the equation of
state.  In addition, the Goldstone nature of the pion, which we have
confirmed explicitly above, allows additional information to be extracted from
plots of $M_\pi^2 $ vs. $<\bar\psi\psi>^2$.
We refer the interested reader to Ref.[18] for
derivations.  Here it is more appropriate to quote results and apply them
to our simulation data.  It is convenient to express the results in terms
of $\eta$, the anomalous dimension of $<\bar\psi\psi>$.
According to hyperscaling the
anomalous dimension $\eta$ is related to the critical indices $\beta_{mag}$
of the order
parameter and $\nu$ of the correlation length by $\beta_{mag}/\nu
= 1 + \eta/2$.  Ref.[18]
predicts the squared pion mass' dependence on the square of the chiral
order parameter within the scaling region for fixed coupling $\beta$.  Consider
the broken symmetry phase $\beta < \beta_c$, the critical point $\beta =
\beta_c$ and the symmetric phase $\beta < \beta_c$ in turn,

$$
\beta < \beta_c:\,\,\,\,\,\,\,\, M_\pi^2 \sim <\bar\psi\psi>^2-
<\bar\psi\psi>_0
\eqno(5.8a)
$$
$$
\beta = \beta_c:\,\,\,\,\,\,\,\, M_\pi^2 \sim (<\bar\psi\psi>^2)^{1/(1+\eta/2)}
\eqno(5.8b)
$$
$$
\beta >\beta_c:\,\,\,\,\,\,\,\, M_\pi^2 \sim (<\bar\psi\psi>^2)^{1/(1+2/\eta)}
\eqno(5.8c)
$$
So, Eq.(5.8a) states that $M_\pi^2  $ should vanish linearly as $m\to 0$ and
$<\bar\psi\psi>^2\to<\bar\psi\psi>_0^2\not=  0$.
Eq.(5.8b) states that at the critical
point $M_\pi^2  $ vs $<\bar\psi\psi>^2$
should be concave downward and provide a
measurement of the anomalous dimension $\eta$.  Recall that a physically
meaningful theory must have
$\eta >0$, so $1/(1 + \eta/2)< 1$.  Finally, Eq.(5.8c)
states that the curves in
the symmetric phase will also be concave downward and will be sensitive to
$\eta$.  Clearly, Eq.(5.8c) does not apply in the case $\eta = 0$.
The more general
formulas of Ref.[18] show that in a mean field scenario where $\eta = 0$
that $M_\pi^2  $
should vary as $C(\beta - \beta_c) + <\bar\psi\psi>^2$ in the symmetric phase.

In summary, we learn from Eq.(5.8) that plots of $M_\pi^2$ vs.
$<\bar\psi\psi>^2$ are sensitive probes into the dynamics of the critical
point.
For example, if mean field theory applies then the $M_\pi^2  $ curves should
be linear functions of $<\bar\psi\psi>^2$ for all coupling and the curve which
passes through the origin picks out the critical point uniquely.  If an
interacting field theory describes the transition, then the $M_\pi^2$
curves should be linear only in the broken symmetry phase and otherwise
have downward curvature determined by the anomalous dimension $\eta$.

In Fig.20 we present our $M_\pi^2 $ and $<\bar\psi\psi>^2$ data in this format.
Clearly the curves are linear for $\beta\leq 0.220$ and give a non-zero chiral
condensate in the chiral limit $M_\pi^2 \to 0$. The functional dependence of
the $\beta =0.225$ and 0.230 curves is not decisively predicted, while the
$\beta = 0.240$ and 0.250 curves are concave downward.  We obtain
an estimate of $\beta_c =
0.225 - 0.230$ in good agreement with our other approaches and, although our
data is not good enough to yield a useful value for $\eta$, we obtain
qualitative evidence that the critical point is not described by a mean
field theory.
\vskip5truemm

\noindent
e.  Mass Ratios and Chiral Symmetry Restoration at Weak Coupling

One of our aims in these simulations is the determination of meson mass
ratios in the chiral limit.  As shown in our discussion of
$M_\pi^2 /M_\sigma^2$  these quantities can indicate whether the chiral
critical
point in trivial and teach us about the realization of chiral symmetry in
the two phases.  Unfortunately our rho and $A_1$ data are not quantitative
enough to use in a similar fashion.  Therefore, we shall only consider $m =
0.01$ data here and look for trends which we believe should survive the
$m\to 0$ extrapolation.

First consider the ratios $M_\pi^2 /M_\sigma^2$  and
$M_\pi^2 /M_\rho^2$  at variable coupling $\beta$.
In Fig.21 we present the results which indicate that
both ratios have non-zero values in the weak coupling phase.  In fact
$M_\pi^2 /M_\sigma^2$ rapidly increases to 0.915(15) at $\beta = 0.26$.
We expect
$M_\pi^2 /M_\sigma^2=1$ in a phase where chiral symmetry is not spontaneously
broken, so this is a welcome result.  Looking back at Fig.17 we see
$R(m,\beta)$ rapidly rising toward unity as $m\to 0$ for $\beta$
values of 0.24 and 0.25, so
this particular result looks rather firm.  Note from Fig.21 that
$M_\pi^2 /M_\rho^2$ appears to be non-zero in the
weak coupling phase but its particular value is not related to symmetry
considerations.

Next, consider the ratio $M_\rho^2 /M_{A_1}^2$ at $m = 0.01$ for various
couplings as plotted in Fig.22.  At weak coupling, $\beta =0.26$ say, the ratio
is consistent with unity, which is the expected value in a phase where
chiral symmetry is an unbroken symmetry,

$$
{ {M_{A_1}^2}\over{M_\rho^2 } }=0.96(8)\,\,\,\,\,\,\,\, ({\rm weak
\,\,\, coupling})
\eqno(5.9)
$$
As we pass to the strong coupling phase this ratio has certainly decreased,
but its precise value is obscured by large error bars.  The data suggests
considerable splitting in the strong coupling phase,

$$
{ {M_{A_1}^2}\over{M_\rho^2 } }\approx 1.35 \,\,\,\,\,\,\,\, ({\rm broken
\,\,\, phase})
\eqno(5.10)
$$
but we hesitate to place an error bar on Eq.(5.10).

The ratio $M_\sigma^2/M_\rho^2$
at $m =0.01$ varies much less dramatically as $\beta$
is varied.  In the vicinity of the critical point and at weak coupling,
Fig.23 shows that

$$
{ {M_\sigma^2}\over{M_\rho^2} }\approx 0.35
\eqno(5.11)
$$
A similar impression comes from Fig.24 where $M_\sigma^2/M_\rho^2$
is plotted against
$M_\pi^2/M_\rho^2$ but the figure is obscured by large error bars.
\vskip10truemm

\centerline{\bf 6. Conclusions and Prospects}
\vskip3truemm

Instead of summarizing our results, we shall end with a few words about
future work.  It should be clear to the reader, however, that considerable
progress has been made in understanding the two flavor version of
noncompact lattice QED and its chiral transition.  That transition appears
to be characterized by non-mean field exponents and monopoles probably play
an important role in its existence.  The use of the Equation of State and
spectroscopy data appears to be a particularly productive approach towards
establishing its physical characteristics.

Obviously we must confirm these results on larger lattices and push our
finite size scaling studies further before solid claims can be made about
the continuum limit of the lattice theory.  Spectroscopy studies using wall
sources should be made on asymmetric lattices with ten times the statistics
accumulated here.  A serious simulation on $64\times 32^3$ lattices seems
essential.

The physical basis of the chiral transition at $\beta_c =0.225(5)$ must be
clarified.  We need to obtain a quantitative understanding of the
importance of monopoles at the transition.  We have already begun a project
in that direction.  By cooling lattice configurations already generated
here, we can produce configurations with the monopoles intact, but the
photons removed.  Is $<\bar\psi\psi>$
the same in these cooled configurations for
$\beta< 0.225$?  Additional calculations of this variety may clarify the
content
of the lattice theory's critical Action.  Interesting work in this area has
already appeared [19].  What is the fate of the fermion and the photon in
the broken symmetry phase?  Our attempts to do direct measurements of these
propagators were not informative due to statistical noise.  In the future
we hope to address these crucial questions using improved wavefunction
techniques borrowed from lattice QCD spectrum calculations.
\vskip1truecm

\centerline{\bf Acknowledgement}
\vskip3truemm

The computer simulation data presented and analyzed here required 3000
hours on one 8K quad of the Connection Machine (CM-2) of NCSA.  Simulations
were also done using the CM-2 at NPAC and PSC, while the ST-100 of Argonne
National Laboratory contributed some data before its untimely demise.  The
work of J. B. Kogut is supported in part by the National Science
Foundation,
NSF-PHY97-00148.  D. K. Sinclair is supported by DOE contract
W-31-109-ENG-38.
\vfill\eject

\noindent
{\bf References}
\vskip3truemm

\noindent
 1. E. Dagotto and J. B. Kogut, Phys. Rev. Lett. {\bf 59}, 617 (1987).

\noindent
2. E. Dagotto, A. Koci\' c and J. B. Kogut, Phys. Rev. Lett. {\bf 60},
772 (1988); {\bf 61}, 2416 (1988).

\noindent
3. S. J. Hands, R. L. Renken, A. Koci\' c, J. B. Kogut, D. K. Sinclair and
K. C. Wang, Phys. Lett. {\bf B261}, 294 (1991).

\noindent
4. L. D. Landau and I. Ya. Pomeranchuk, Dokl. Akad. Nauk {\bf 102}, 489 (1955).

\noindent
5. T. DeGrand and D. Toussaint, Phys. Rev. {\bf D22}, 2478 (1980);
J. Cardy, Nucl. Phys. {\bf B170}[FS1], 369 (1980);
A. Ukawa, P. windey and A. Guth, Phys. Rev. {\bf D21}, 1013 (1980).

\noindent
6. P.A.M. Dirac, Phys. Rev. {\bf 74}, 817 (1948).

\noindent
7. W. Bardeen, T. Leung and S. Love, Nucl. Phys. {\bf B273}, 649 (1986);
W. Bardeen, S. Love and V. Miransky, Phys. Rev. {\bf D42}, 3514 (1990).

\noindent
8. A. Koci\' c, S. Hands, E. Dagotto and J. B. Kogut,
Nucl. Phys. {\bf B347}, 217 (1990).

\noindent
9. E. Dagotto and J. B. Kogut, Nucl. Phys. {\bf B295}[FS21], 123 (1988);
E. Dagotto, A. Koci\' c and J. B. Kogut, ibid. {\bf B317}, 271 (1989);
{\bf B331}, 500 (1990).

\noindent
10. S. Duane, Nucl. Phys. {\bf B257}[FS14], 612 (1985).

\noindent
11. C. Itzykson and J.-M. Drouffe, {\it Statistical Field Theory}, Cambridge
University Press, Cambridge (1989).

\noindent
12. S. Hands, E. Dagotto and J.B. Kogut, Nucl. Phys. {\bf B333}, 551 (1990);

\noindent
13. S. Hands and R. Wensley, Phys. Rev. Lett. {\bf 63}, 2169 (1989).

\noindent
14. S. Hands, A. Koci\' c and J. B. Kogut, ILL-(TH)-92-6 (to apper in Phys.
Lett.).

\noindent
15. V. Grosch, K. Jansen, J. Jers\' ak, C. Lang, T. Neuhaus and C. Rebbi, Phys.
Lett. {\bf B162}, 171 (1985).

\noindent
16. S. Kirkpatrick, Phys. Rev. Lett. {\bf 36}, 69 (1976).

\noindent
17. S. Gottlieb, W. Liu, R. L. Renken, R. Sugar and D. Toussaint, Phys.
Rev. {\bf D38}, 2245 (1988).

\noindent
18. A. Koci\' c, J. B. Kogut and M.-P. Lombardo, " Universal properties of
Chiral Symmetry Breaking" Illinois - CERN preprint;
A. Koci\' c, Phys. Lett. {\bf B281}, 309 (1992).

\noindent
19. V. Azcoiti, G. DiCarlo and A. F. Grillo, DFTUZ.91/34 and references
therein.
\vfill\eject

\baselineskip=20pt
\parskip=5pt plus 1pt
\parindent=15pt
\centerline{\bf Table 1}
\vskip 0.5 truecm
\centerline{
Average plaquette data at various bare fermion masses $m$ and couplings
$\beta=1/e^2$ on a $16^4$ lattice.}
\vskip 1 truecm
$$\vbox{\settabs\+\qquad.225\qquad&\qquad1.0230(4)\qquad&\qquad1.0266(4)
\qquad&\qquad1.0290(3)\qquad&\qquad1.0302(3)\qquad&\cr
\+\hfill$\beta$\hfill&\hfill$m=.005$\hfill&\hfill$m=.010$\hfill&
\hfill$m=.020$\hfill&\qquad$m=.030$\qquad&\cr
\bigskip
\+\hfill.30\hfill&\hfill--\hfill&\hfill--\hfill&\hfill0.7803(2)
\hfill&\hfill0.7815(2)\hfill&\cr\smallskip
\+\hfill.28\hfill&\hfill--\hfill&\hfill--\hfill&\hfill0.8332(2)
\hfill&\hfill0.8338(2)\hfill&\cr\smallskip
\+\hfill.27\hfill&\hfill--\hfill&\hfill--\hfill&\hfill0.8621(3)
\hfill&\hfill0.8625(3)\hfill&\cr\smallskip
\+\hfill.26\hfill&\hfill--\hfill&\hfill0.8922(3)\hfill&\hfill0.8943(2)
\hfill&\hfill0.8949(3)\hfill&\cr\smallskip
\+\hfill.25\hfill&\hfill--\hfill&\hfill0.9264(4)\hfill&\hfill0.9283(2)
\hfill&\hfill0.9296(3)\hfill&\cr\smallskip
\+\hfill.24\hfill&\hfill--\hfill&\hfill0.9631(4)\hfill&\hfill0.9651(3)
\hfill&\hfill0.9662(3)\hfill&\cr\smallskip
\+\hfill.235\hfill&\hfill--\hfill&\hfill--\hfill&\hfill0.9858(3)
\hfill&\hfill0.9869(3)\hfill&\cr\smallskip
\+\hfill.23\hfill&\hfill1.0025(2)\hfill&\hfill1.0043(4)\hfill&\hfill1.0070(3)
\hfill&\hfill1.0076(2)\hfill&\cr\smallskip
\+\qquad.225\qquad&\qquad1.0230(4)\qquad&\qquad1.0266(4)
\qquad&\qquad1.0290(3)\qquad&\qquad1.0302(3)\qquad&\cr\smallskip
\+\hfill.22\hfill&\hfill1.0465(3)\hfill&\hfill1.0489(4)\hfill&\hfill1.0521(3)
\hfill&\hfill1.0539(3)\hfill&\cr\smallskip
\+\hfill.215\hfill&\hfill--\hfill&\hfill--\hfill&\hfill1.0772(4)
\hfill&\hfill1.0783(3)\hfill&\cr\smallskip
\+\hfill.21\hfill&\hfill--\hfill&\hfill1.1011(4)\hfill&\hfill1.1049(5)
\hfill&\hfill1.1047(3)\hfill&\cr\smallskip
\+\hfill.20\hfill&\hfill--\hfill&\hfill1.1607(9)\hfill&\hfill1.1635(5)
\hfill&\hfill1.1628(4)\hfill&\cr\smallskip
\+\hfill.19\hfill&\hfill--\hfill&\hfill--\hfill&\hfill1.2300(5)
\hfill&\hfill1.2276(4)\hfill&\cr\smallskip
\+\hfill.18\hfill&\hfill--\hfill&\hfill--\hfill&\hfill1.3028(7)
\hfill&\hfill1.2988(4)\hfill&\cr\smallskip
\+\hfill.17\hfill&\hfill--\hfill&\hfill--\hfill&\hfill1.3883(6)
\hfill&\hfill1.3818(5)\hfill&\cr\smallskip
\+\hfill.16\hfill&\hfill--\hfill&\hfill--\hfill&\hfill1.4821(7)
\hfill&\hfill1.4757(4)\hfill&\cr\smallskip
\+\hfill.15\hfill&\hfill--\hfill&\hfill--\hfill&\hfill1.5873(9)
\hfill&\hfill1.5812(4)\hfill&\cr}$$
\vfill\eject
\centerline{\bf Table 2}
\vskip 0.5 truecm
\centerline{
$<\bar\psi\psi>$ data at various bare fermion masses $m$ and couplings
$\beta=1/e^2$ on a $16^4$ lattice.}
\vskip 1 truecm
$$\vbox{\settabs\+\qquad.225\qquad&\qquad.0943(8)\qquad&\qquad.2585(14)
\qquad&\qquad.2200(11)\qquad&\qquad.2119(3)\qquad&\cr
\+\hfill$\beta$\hfill&\hfill$m=.005$\hfill&\hfill$m=.010$\hfill&
\hfill$m=.020$\hfill&\qquad$m=.030$\qquad&\cr
\bigskip
\+\hfill.30\hfill&\hfill--\hfill&\hfill--\hfill&\hfill.0567(1)
\hfill&\hfill.0821(1)\hfill&\cr\smallskip
\+\hfill.28\hfill&\hfill--\hfill&\hfill--\hfill&\hfill.0697(1)
\hfill&\hfill.0988(1)\hfill&\cr\smallskip
\+\hfill.27\hfill&\hfill--\hfill&\hfill--\hfill&\hfill.0793(2)
\hfill&\hfill.1111(2)\hfill&\cr\smallskip
\+\hfill.26\hfill&\hfill--\hfill&\hfill.0499(2)\hfill&\hfill.0924(2)
\hfill&\hfill.1265(2)\hfill&\cr\smallskip
\+\hfill.25\hfill&\hfill--\hfill&\hfill.0630(4)\hfill&\hfill.1098(3)
\hfill&\hfill.1452(3)\hfill&\cr\smallskip
\+\hfill.24\hfill&\hfill--\hfill&\hfill.0824(5)\hfill&\hfill.1331(3)
\hfill&\hfill.1675(3)\hfill&\cr\smallskip
\+\hfill.235\hfill&\hfill--\hfill&\hfill--\hfill&\hfill.1483(4)
\hfill&\hfill.1812(2)\hfill&\cr\smallskip
\+\hfill.23\hfill&\hfill.0764(9)\hfill&\hfill.1135(8)\hfill&\hfill.1636(6)
\hfill&\hfill.1955(2)\hfill&\cr\smallskip
\+\qquad.225\qquad&\qquad.0943(8)\qquad&\hfill.1340(7)
\hfill&\hfill.1809(5)\hfill&\qquad.2119(3)\qquad&\cr\smallskip
\+\hfill.22\hfill&\hfill.1225(9)\hfill&\hfill.1539(6)\hfill&\hfill.1980(6)
\hfill&\hfill.2288(4)\hfill&\cr\smallskip
\+\hfill.215\hfill&\hfill--\hfill&\hfill--\hfill&\qquad.2200(11)
\qquad&\hfill.2462(2)\hfill&\cr\smallskip
\+\hfill.21\hfill&\hfill--\hfill&\hfill.2056(9)\hfill&\hfill.2395(9)
\hfill&\hfill.2638(3)\hfill&\cr\smallskip
\+\hfill.20\hfill&\hfill--\hfill&\qquad.2585(14)\qquad&\hfill.2827(8)
\hfill&\hfill.3019(4)\hfill&\cr\smallskip
\+\hfill.19\hfill&\hfill--\hfill&\hfill--\hfill&\hfill.3261(7)
\hfill&\hfill.3409(4)\hfill&\cr\smallskip
\+\hfill.18\hfill&\hfill--\hfill&\hfill--\hfill&\hfill.3678(11)
\hfill&\hfill.3772(3)\hfill&\cr\smallskip
\+\hfill.17\hfill&\hfill--\hfill&\hfill--\hfill&\hfill.4085(10)
\hfill&\hfill.4138(3)\hfill&\cr\smallskip
\+\hfill.16\hfill&\hfill--\hfill&\hfill--\hfill&\hfill.4427(7)
\hfill&\hfill.4481(2)\hfill&\cr\smallskip
\+\hfill.15\hfill&\hfill--\hfill&\hfill--\hfill&\hfill.4774(9)
\hfill&\hfill.4792(2)\hfill&\cr}$$
\vfill\eject
\centerline{\bf Table 3}
\vskip 0.5 truecm
\noindent
Peak of the monopole susceptibility and the order parameter $M=n_{max}/n_{tot}$
at $\beta_c=.225$ on $L^4$ lattices with $L$ ranging from 10 to 18.
\vskip 1 truecm
$$\vbox{\settabs\+\qquad18\qquad&\qquad190.4(8.8)\qquad&\qquad.1630(53)
\qquad&\cr
\+\hfill$L$\hfill&\hfill$\chi_{max}$\hfill&\hfill$M$\hfill&\cr\bigskip
\+\hfill10\hfill&\hfill53.3(2.4)\hfill&\hfill.222(12)\hfill&\cr\smallskip
\+\hfill12\hfill&\hfill78.3(4.1)\hfill&\hfill.212(13)\hfill&\cr\smallskip
\+\hfill14\hfill&\hfill108.7(6.2)\hfill&\hfill.193(14)\hfill&\cr\smallskip
\+\hfill16\hfill&\hfill143.1(5.2)\hfill&\hfill.177(7)\hfill&\cr\smallskip
\+\qquad18\qquad&\qquad190.4(8.8)\qquad&\qquad.1630(53)\hfill&\cr}$$
\vfill\eject
\centerline{\bf Table 4}
\vskip 1 truecm
\centerline{
$m^2_\pi$ values for various couplings $\beta$ and bare fermion masses $m$.}
\vskip 1 truecm
$$\vbox{
\settabs\+\quad$m\backslash\beta$\quad&0.229(3)\quad&0.246(3)
\quad&0.266(3)\quad&0.264(12)\quad&0.279(14)\quad&0.272(45)
\quad&0.287(27)\quad&0.293(25)\quad&\cr
\+\quad$m\backslash\beta$\quad&\quad.20\hfill&\quad.21\hfill&\quad.22
\hfill&\quad.225\hfill&\quad.23\hfill&\quad.24\hfill&\quad.25\hfill&
\quad.26\hfill&\cr\bigskip
\+\hfill.03\hfill&0.229(3)\quad&0.246(3)\quad&0.266(3)\quad&
0.264(12)\quad&0.279(14)\quad&0.272(45)\quad&0.287(27)\quad&
0.293(25)\quad&\cr\smallskip
\+\hfill.02\hfill&0.159(2)\hfill&0.176(2)\hfill&0.197(2)\hfill&
0.212(4)\hfill&0.207(4)\hfill&0.237(5)\hfill&0.254(5)\hfill
&0.296(6)\hfill&\cr\smallskip
\+\hfill.01\hfill&0.083(2)\hfill&0.096(3)\hfill&0.114(4)\hfill&
0.124(3)\hfill&0.137(3)\hfill&0.171(9)\hfill&0.214(5)\hfill
&0.250(5)\hfill&\cr\smallskip
\+\hfill.005\hfill&\quad--\hfill&\quad--\hfill&0.069(3)\hfill&
0.082(2)\hfill&0.095(2)\hfill&\quad--\hfill&\quad--\hfill
&\quad--\hfill&\cr}$$
\vfill
\centerline{\bf Table 5}
\centerline{
$m^2_\sigma$ values for various couplings $\beta$ and bare fermion masses $m$.
}
\vskip 1 truecm
$$\vbox
{\settabs\+\quad $m\backslash\beta$\quad &1.051(49)\quad&0.992(32)
\quad&0.709(22)\quad&0.681(16)\quad&0.662(11)\quad&0.567(17)
\quad&0.505(10)\quad&0.522(8)\quad&\cr
\+\quad $m\backslash\beta$\quad&\quad.20\hfill&\quad.21\hfill&\quad.22
\hfill&\quad.225\hfill&\quad.23\hfill&\quad.24\hfill&\quad.25\hfill&\quad
.26\hfill&\cr\bigskip
\+\hfill.03\hfill&1.051(49)\quad&0.992(32)\quad&0.709(22)\quad&
0.681(16)\quad&0.662(11)\quad&0.567(17)\quad&0.505(10)\quad&
0.522(8)\quad&\cr\smallskip
\+\hfill.02\hfill&1.136(78)\hfill&0.716(48)\hfill&0.570(16)
\hfill&
0.520(20)\hfill&0.455(13)\hfill&0.452(15)\hfill&0.380(8)
\hfill
&0.386(7)\hfill&\cr\smallskip
\+\hfill.01\hfill&0.610(87)\hfill&0.475(22)\hfill&0.358(12)
\hfill&
0.339(11)\hfill&0.286(9)\hfill&0.246(7)\hfill&0.252(5)\hfill
&0.273(5)\hfill&\cr\smallskip
\+\hfill.005\hfill&\quad--\hfill&\quad--\hfill&0.233(11)\hfill&
0.205(6)\hfill&0.173(6)\hfill&\quad--\hfill&\quad--\hfill
&\quad--\hfill&\cr}$$
\vfill\eject
\centerline{\bf Table 6}
\centerline{
$m^2_\rho$ values for various couplings $\beta$ and bare fermion masses $m$.}
\vskip 1 truecm
$$\vbox
{\settabs\+\quad$m\backslash\beta$\quad&2.173(15)\quad&1.399(29)
\quad&1.409(22)\quad&1.681(66)\quad&1.662(81)\quad&
1.567(17)
\quad&0.505(10)\quad&0.522(78)\quad&\cr
\+\quad$m\backslash\beta$\quad&\quad.20\hfill&\quad.21\hfill&\quad.22
\hfill&\quad.225\hfill&\quad.23\hfill&\quad.24\hfill&\quad.25\hfill&\quad
.26\hfill&\cr\bigskip
\+\hfill.03\hfill&2.17(15)\hfill&1.073(93)\hfill&0.412(54)
\hfill&
1.36(11)\quad&1.24(11)\quad&1.195(71)\quad&0.789(34)\quad&
0.837(35)\quad&\cr\smallskip
\+\hfill.02\hfill&0.83(32)\hfill&1.25(91)\quad&\quad--
\hfill&
1.17(40)\hfill&\quad--\hfill&\quad--\hfill&0.645(52)
\hfill
&0.564(99)\hfill&\cr\smallskip
\+\hfill.01\hfill&1.15(25)\hfill&1.11(18)\hfill&1.03(17)
\quad&
0.956(88)\hfill&0.817(75)\hfill&0.776(62)\hfill&0.767(55)
\hfill
&0.663(33)\hfill&\cr\smallskip
\+\hfill.005\hfill&\quad--\hfill&\quad--\hfill&0.56(19)\hfill&
0.98(32)\hfill&0.57(11)\hfill&\quad--\hfill&\quad--\hfill
&\quad--\hfill&\cr}$$
\vfill
\centerline{\bf Table 7}
\centerline{
$m^2_A$ values for various couplings $\beta$ and bare fermion masses $m$.}
\vskip 1 truecm
$$\vbox
{\settabs\+\qquad$m\backslash\beta$\qquad&2.16(1.15)\quad&
1.99(22)
\quad&1.79(22)\quad&1.68(16)\quad&1.66(81)\quad&
1.567(71)
\quad&1.505(10)\quad&1.522(78)\quad&\cr
\+\quad$m\backslash\beta$\quad&\quad.20\hfill&\quad.21\hfill&\quad.22
\hfill&\quad.225\hfill&\quad.23\hfill&\quad.24\hfill&\quad.25\hfill&\quad
.26\hfill&\cr\bigskip
\+\hfill.03\hfill&2.19(1.23)\quad&\quad--\hfill&1.19(23)\quad&
1.69(38)\quad&1.61(29)\quad&1.49(23)\quad&1.113(97)\quad&
1.201(85)\quad&\cr\smallskip
\+\hfill.02\hfill&1.80(1.23)\hfill&1.99(80)\hfill&
1.46(28)\hfill&
1.70(21)\hfill&1.85(19)\hfill&1.43(15)\hfill&
1.580(97)
\hfill
&1.160(72)\hfill&\cr\smallskip
\+\hfill.01\hfill&0.87(48)\hfill&0.74(28)\hfill&1.39(35)
\hfill&
1.08(15)\hfill&1.18(14)\hfill&0.920(94)\hfill&
0.808(62)
\hfill
&0.693(35)\hfill&\cr\smallskip
\+\hfill.005\hfill&\quad--\hfill&\quad--\hfill&\quad--\hfill&
\quad--\hfill&\quad--\hfill&\quad--\hfill&\quad--\hfill
&\quad--\hfill&\cr}$$
\vfill\eject

\noindent
{\bf Figure Captions}
\vskip5truemm

1. $<\bar\psi\psi> $ vs. $\beta$ for $m = 0.03$, 0.02, 0.01 and 0.005
for $16^4$ lattice.  The
data is given in Table 1.  The size of the symbols in the figure includes
the statistical error   bars.

\noindent
2. $-\ln <\bar\psi\psi> $ vs. $-\ln \, m$ plots for $\beta = 0.22$, 0.225
and 0.23.

\noindent
3. $-1/\ln(m)$ vs. $-1/\ln <\bar\psi\psi> $ plots for various $\beta$ values.

\noindent
4. Equation of State Eq.(2.1) for $\beta_c = 0.220$,
$\delta = 2.75$, $\beta_{mag} = 0.571$.

\noindent
5. Equation of State Eq.(2.1) for $\beta_c = 0.225$, $\delta = 2.31$,
$\beta_{mag} = 0.763$.

\noindent
6. Equation of State Eq.(2.1) for $\beta_c = 0.230$, $\delta = 1.89$,
$\beta_{mag} = 1.124$.

\noindent
7. Monopole susceptibility $\chi$ vs. $\beta$ on a $16^4$
lattice and bare fermion masses 0.03,   0.02 and 0.01.

\noindent
8. $\ln \chi_{max}$ vs. $\ln \, L$ plotted from the data in Table 2.
The lattices range
in linear dimension $L = 10$, 12, 14, 16, and 18, and the coupling is
$\beta_c =0.225$.

\noindent
9. a. Histogram of the monopole order parameter at $\beta = 0.24$.

\noindent
b. Histogram of the monopole order parameter at $\beta = 0.23$.

\noindent
c.      Histogram of the monopole order parameter at $\beta = 0.225.$

\noindent
d.      Histogram of the monopole order parameter at $\beta = 0.22.$

\noindent
e.      Histogram of the monopole order parameter at $\beta = 0.21.$

\noindent
f.      Histogram of the monopole order parameter at $\beta = 0.20.$

\noindent
10. $\ln M(\beta_c,L)$ vs. $\ln \, L$ from the data in Table 3 at
$\beta_c = 0.225.$

\noindent
11. The monopole concentration $<\ell>/N_\ell$ (triangles) and
the percolation order   parameter $M$ (circles) on a $16^4$ lattice at
$m = 0.01$.

\noindent
12. A typical pion propagator and its fit, $\beta = 0.225$ and $m = 0.01$.

\noindent
13. A typical sigma propagator and its fit, $\beta = 0.225$ and
$m = 0.01$.

\noindent
14. A typical rho propagator and its fit, $\beta = 0.225$ and $m = 0.01$.

\noindent
15. A typical $A_1$ propagator and its fit, $\beta = 0.225$ and $m = 0.01$.

\noindent
16. $M_\pi^2$ vs. $\beta$ at $m\to 0$.
The plot was obtained by extrapolating the
data in Table 4 to $m = 0$ as discussed in the text.

\noindent
17. $R(m,\beta)$ vs. $m$ for various couplings $\beta$ 0.200 (circles), 0.210
(triangles), 0.220 (inverted triangles), 0.225 (hexagons), 0.230 (dark
circles), 0.240 (dark squares)  and 0.250 (dark triangles).

\noindent
18. The chiral Equation of State plotted in the form Eq.(5.3) with $\beta_c =
0.225$, $\delta = 2.31$ and $\beta_{mag} = 0.763$.
The line is the best linear fit to the universal function $f$.

\noindent
19. $R(m,\beta)$ vs. $m$ from Eq.(5.4a) and (5.6) at $\beta$
values coinciding with Fig.17.

\noindent
20. $M_\pi^2$ vs. $<\bar\psi\psi>^2$ for the data of Tables 2 and 4.

\noindent
21. $M_\pi^2/M_\sigma^2$ and $M_\pi^2/M_\rho^2$ vs. $\beta$ for
fixed $m =0.01$.

\noindent
22. $M_\rho^2/M_{A_1}^2$  vs. $\beta$ for $m = 0.01$.

\noindent
23. $M_\sigma^2/M_\rho^2$  vs. $\beta$ for $m = 0.01$.

\noindent
24. $M_\sigma^2/M_\rho^2$   vs. $M_\pi^2/M_\rho^2$  for $m = 0.01$.
\vfill\eject

\end